# Thermal fluctuations and vortex melting in the classical superconductor $Nb_3Sn$ from high-resolution specific-heat measurements

R. Lortz[1], F. Lin[2], N. Musolino[1], Y. Wang[1], A. Junod[1], B. Rosenstein[2], N. Toyota[3]

[1]Department of Condensed Matter Physics, University of Geneva, 24 Quai Ernest-Ansermet, CH-1211 Geneva 4, Switzerland

[2]National Center for Theoretical Sciences and Electrophysics Department, National Chiao Tung University, Hsinchu 30050, Taiwan

[3]Physics Department, Graduate School of Science, Tohoku University, 980-8571 Sendai, Japan

The range of critical thermal fluctuations in 'classical' bulk superconductors is extremely small and especially in low fields hardly experimentally inaccessible. With a new type of calorimeter we have been able to resolve a small lambda anomaly within a narrow temperature range around the $H_{c2}$ line. We show that the evolution of the anomaly as a function of magnetic field follows scaling laws expected in the presence of critical fluctuations. The lower onset of the fluctuation regime shows many characteristics of a continuous solid-to-liquid transition in the vortex matter. It can be driven into a first-order vortex melting transition by a small AC field which helps the vortex matter to reach equilibrium.

## I. INTRODUCTION

The superconducting transition belongs to the family of second-order phase transitions for which the order parameter is continuous at the critical temperature. Fluctuations typically dominate such transitions. This is certainly true for high-temperature superconductors (HTSCs) [1-9] and layered organic superconductors [10,11] in which fluctuations broaden the superconducting transition in a magnetic field and melt the vortex lattice into a liquid far below the mean-field transition temperature. However, in the case of 'classical' bulk low-$T_c$ superconductors the coherence volumes are large and the transition temperatures low. This strongly narrows the temperature window around $T_c$ in zero magnetic fields in which the effect of fluctuations becomes important down to a size which is generally believed to be experimentally inaccessible. Experimental limits for their observation are sample inhomogeneity and the temperature resolution of standard thermodynamic measurements. Improving sample quality and experimental sensitivity to observe critical fluctuations thus poses a significant challenge to the experimentalist. Nevertheless, some 'universal' features in the field-temperature phase diagram of classical type-II superconductors, e.g. a broadening of the upper-critical field line ($H_{c2}$ or $T_c(H)$) or a peak effect [13-18] show analogies with the HTSCs. New high-resolution calorimetric methods which we developed during years of study on the HTSCs motivated us to investigate whether fluctuation effects and vortex melting might also be responsible in this case. We have chosen the compound $Nb_3Sn$, as high purity single crystals are available [19] and it is still the most important superconductor for applications.

In this paper, we report specific-heat measurements on a homogeneous single crystal of the superconductor $Nb_3Sn$. We observe a small lambda anomaly superimposed on the specific-heat jump at $T_c(H)$. The broadening of the anomaly and the jump as a function of magnetic field follows scaling laws typical in the presence of fluctuations. The lower onset of the fluctuation regime may be interpreted as an 'idealized' upward step related to a continuous vortex melting transition. It can be driven into a first-order transition by a small AC field. Fluctuations are also observed in the sample magnetization which proves that the anomalies are of a thermodynamic origin.

## II. EXPERIMENTS

The sample under investigation is a Nb$_3$Sn single crystal with a $T_c$ width of ~20 mK [19]. The specific heat was measured with a new type of quasi-isothermal calorimeter. A sample platform with a deposited thin-film heater is suspended by a thermopile made of 24 thermocouples acting as a strong heat link to the surrounding thermal bath. The specific heat can be obtained using an AC technique [21] which provides a high sensitivity and density of data points, or with a less sensitive DC heat-flow technique [22]. The latter is helpful to calibrate absolute values and to rule out artifacts due to the AC method. These may arise in the presence thermal hysteresis effects, e.g. due to flux pinning or close to a first-order phase transition. The isothermal magnetization was measured with a commercial Quantum Design MPMS-5 SQUID magnetometer using a scan length of 2 cm in steps of 4 mT.

An overview of the specific heat measured with the AC technique in various fields is given in Fig. 1a. Figure 1b shows details close to $T_c$(0). In zero fields the typical mean-field jump is found at $T_c$. No signature of fluctuations is visible when entering the superconducting phase from the normal state. In a small field > 0.05 T the jump D$C$ is reduced by a factor of 1.11. For high-κ superconductors (κ is the ratio of penetration depth λ to coherence length ξ) a reduction of D$C$ by a factor of 1.16 is expected when entering the Abrikosov phase instead of the Meissner phase [23], close to our value. For μ$_0$H = 0.05 T the transition at the $H_{c1}$ line can be seen as a smooth increase of the specific heat towards the zero-field data upon lowering the temperature below $T_c$($H$). It extends over a broad temperature interval between 17.4 – 17.9 K as the transition line is almost reached tangentially in this temperature sweep experiment. In fields >0.2 T a tiny lambda anomaly, superimposed on the jump, can be resolved. If the field is raised further the anomaly grows until it represents ~10 % of the total jump. The temperature range over which it is visible is enlarged simultaneously with the jump upon increasing the field. To exclude artifacts originating from dissipative effects of the vortex matter in presence of flux pinning when using the AC technique, we compared the data for a few fields with measurements performed in the DC mode. The shape of the anomalies does however not depend on the method used (inset of Fig. 1a). Furthermore, the same anomaly is found in field-sweep experiments [17]. This rules out irreversible effects as its origin. Fig. 2a shows isothermal magnetization measurements. A deviation from linear behavior just below the kink at $H_{c2}$ indicates the onset of fluctuations. The effect appears more clearly in the derivative $dM/dH$ (Fig. 2b) where a similar lambda anomaly to that seen in the specific heat appears superimposed on a jump. Together these measurements prove that the observed anomaly is a true thermodynamic feature.

## III. DISCUSSION

### A. Thermal fluctuations

A measure of the width of the critical regime is the Ginzburg number $Gi=0.5(k_BT_c)^2/(H_c^2(0)\xi_0^3)^2$; $H_c$(0) is the thermodynamic critical field at $T=0$ and $\xi_0$ the isotropic Ginsburg-Landau coherence length. While $Gi$ is enhanced in HTSCs to ~$10^{-1}$-$10^{-3}$, small values are found in bulk classical superconductors, e.g. $10^{-10}$ for Nb. The Ginzburg temperature $t_G=Gi \cdot T_c$ determines the temperature range around $T_c$ where fluctuations in the specific heat are of the same order of magnitude as the mean-field jump. Using $T_c$=18 K, $H_c$(0)=5200 Oe and $\xi_0$ = 36 Å [12], we obtain $t_G$ =$10^{-5}$ - $10^{-4}$ K. Contributions of a few % of the jump might therefore be observable in a range of $10^{-2}$ K around $T_c$. We observe in the 0.25 T data that the anomaly represents 3% of the specific-heat jump and it appears over a temperature range $\sim 10^{-2}$ K. For comparison, $t_G$ for the HTSC YBa$_2$Cu$_3$O$_7$ (YBCO) is of the order of 1 K, while fluctuation contributions have been observed up to 30 K above $T_c$ [4,5]. Furthermore, $t_G$ increases in a field due to a reduction of the effective dimensionality arising from the confinement of the excitations to a few low Landau orbits [24]. This lead to the observation of fluctuations in some classical superconductors, especially in Nb, in rather high fields close to $H_{c2}(0)$ which have been interpreted in terms of 'Lowest Landau Level' (LLL) fluctuations [25,26]. In the present work the fluctuations can be traced down to very low fields where they express themselves in a rather sharp lambda anomaly.

Outside the temperature range defined by $t_G$, the fluctuations in the low-field limit have been described either within the Gaussian or the 3d-XY model including correction terms [1,27]. The temperature range of the fluctuations is too small here to extract any critical exponents. Information can be obtained from the broadening and the increasing width of the fluctuation regime in magnetic fields. Both follow scaling laws as we will show below. In some HTSCs, the scaling of data taken in different fields has been successfully used to investigate fluctuations [2,3,5-7]. The effect of a magnetic field on fluctuations is to induce a length scale which reduces the effective dimensionality. This magnetic length $l=(\Phi_0/B)^{1/2}$ is related to the vortex-vortex distance in type-II superconductors [5,28]. In the critical regime, the correlation length $\xi$ diverges upon approaching $T_c$ following a power law of the form $\xi \sim \xi_0 (1-t)^{-\nu}$, $t=T/T_c(0)$, $\nu=0.67$ (3d-XY). The presence of a magnetic length will cut off this divergence within a certain temperature window. A finite-size effect is thus responsible for the broadening of $T_c$ in fields. Scaling involves normalizing data taken in different fields by the ratio of $\xi$ to the magnetic length. If universality holds scaled data should merge on a common curve. 3d-XY scaling was observed e.g. for YBCO up to fields of $\sim$10 T [5] although the temperature range was clearly larger than $t_G$. In higher fields the quasiparticles are confined to low Landau levels and the scaling model for 3d-LLL fluctuations should be applicable [6,8,26]. We tested both models on Nb$_3$Sn. The 14 T data above 10 K represents the specific heat of the normal state and was used to subtract the phonon contribution. In Fig. 3a we started by 3d-XY scaling the data in low fields. As the fluctuations contribution is small, we had to normalize the specific-heat jump and consider a field-dependent $T_c(H)$. The data can then be merged when plotted as $C/\Delta CH^{\alpha/2\nu}$ versus $[T/T_c(H)-1]^{-1/2\nu}$, which is the proper scaling for the 3d-XY model ($\alpha \cong -0.001$) [1]. The scaling becomes worse around 4 T and fails completely in higher fields. Fig. 3c shows a similar 3d-XY scaling plot for $dM(H)/dH$. The curves (in the low-field 3d-XY limit) merge perfectly and prove that the signature of fluctuations is a true thermodynamic effect. In Fig. 3b we plotted the same data as $C/\Delta C$ versus $[T-T_c(H)](HT)^{-2/3}$ which is the proper normalization for the 3d-LLL model. The scaling holds in fields above 4 T in the region of the broadened specific-heat jump, while the curves measured in lower fields deviate increasingly. To compare the data more quantitatively with the 3d-LLL model in which pinning is neglected we performed a clean 3d-LLL fit [9] with help of the LLL scaling parameter: $a_T=-(2/Gi)^{1/3}[t-T_c(H)/T_c(0)](ht)^{-2/3}$ [25], where $h=H/H_{c2}(0)$. The second critical field line is accurately approximated by: $H_{c2}(T)=c_1[T-T_c(0)]-c_2[T-T_c(0)]^2$, $c_1=-2.05$ and $c_2=-0.036$. Coefficients of the Ginsburg-Landau model such as the effective mass $m^*$ which weakly depend on temperature were extracted from the zero-field specific-heat data. The so obtained theoretical curve is included in Fig. 3b and represents a scaling function. The data at high fields can not only be merged, but also mapped onto the scaling function in the upper temperature range. In fields lower than 4 T the broadening of the curves is overestimated by the 3d-LLL model, indicating contributions from quasiparticles occupying higher Landau orbits and a crossover into the low-field 3d-XY region. According to the theoretical fit a solid vortex phase is the ground state for $a_T < -9.5$ and a liquid phase for $a_T > 9.5$. A vortex melting transition should thus occur at $a_T = -9.5$ [9,29]. It is manifested by a spike as included in Fig. 3b, which was however not observed in the experimental data. The scaling of the experimental data is limited to the temperature range above this theoretical melting temperature. Below a smooth upturn of the specific heat is found instead and will be discussed later. We note that the exponents in the scaling variables of the 3d-XY and the 3d-LLL model are very similar and it is hard to determine from scaling alone which model is more suitable for describing the fluctuations, as has also been observed in HTSCs [2,3]. Nevertheless neither model can be used alone to describe the scaling over the whole range of fields which we take as an indication for a crossover from 3d-XY to 3d-LLL fluctuations in high fields.

The fluctuations disappear in low fields in the vicinity of the Meissner phase. For this classical superconductor with a large coherence volume vortex degrees of freedom are likely to be important to destabilize the phase of the superconducting condensate. The fluctuation anomaly might thus be related to a transition into a liquid vortex phase. Evidence for vortex melting has indeed been found in the experimental data, as discussed below.

### B. Vortex melting

Below 15 K, irreversibility due to enhanced flux pinning close to the $H_{c2}$ line gives rise to a sharp magnetization peak effect (Fig. 3 inset) [17,18]. Such peak effects have been interpreted as being due to an underlying solid-to-liquid transition in the vortex matter [30]. The onset of the fluctuation anomaly in the specific heat when approaching $T_c(H)$ from below may be interpreted as an 'idealized' upward step, similar to the second-order vortex melting observed in some YBCO samples [31]. It has also been shown that a small AC field can help the vortices to reach equilibrium in the presence of flux pinning [32,33]. For this reason we installed a coil below the sample platform of our calorimeter and repeated measurements with a superimposed AC field of ~10 G, 1 kHz applied parallel to the DC field. Starting from 3 T the shape of the fluctuation anomaly changes: The upward step is shifted somewhat to lower temperatures and becomes clearly sharper (Fig. 4a). In 5 T a small spike replaces the smooth upturn in the specific heat which grows further in 6 T which is the highest field in which we were able to perform this experiment. The AC coil which is thermally anchored to the thermal bath dissipates too much heat to stabilize temperatures lower than 13.5 K. In Fig 4b we plotted the difference in the curves with and without AC field to show the anomaly more clearly.

The spike in fields above 4 T most probably arises from the latent heat of a first-order vortex-melting transition. Self-heating effects due to screening currents induced by the AC field can be ruled out as an origin of the spike. The produced heat would flow out of the sample to the thermal bath (negative contribution to the specific heat), while heat is absorbed from the thermal bath by the latent heat of a first-order transition which results in a positive contribution. AC specific-heat methods (as also some relaxation methods) may underestimate latent heats of first-order transitions if the temperature hysteresis is larger than the amplitude of the temperature modulation of the experiment. This was e.g. demonstrated on the first-order vortex-melting transition in YBCO [34]. The AC method with a small temperature modulation (~1 mK) was however necessary to have sufficiently temperature resolution to resolve the spikes at the melting temperature ($T_m$) without broadening it. To measure the true value of the latent heat which can be compared to that observed in YBCO, we measured the DC heat flow through the thermopile during a field sweep in presence of the small AC field at a fixed temperature (13.5 K). The related quantity is the isothermal magnetocaloric effect $M_T(H) = (dQ/dB)_{T=const}$ (see Ref. 17 for more details). Also in this quantity, which is closely related to the specific heat, a spike is observed at ~7 T (Inset of Fig. 4b). From the area below the spike we can calculate the latent heat absorbed at the transition from the solid to the liquid phase. We find a value of $\Delta S$=2 $10^{-3}$ ± 5 $10^{-4}$ J gat$^{-1}$ K$^{-1}$ = 0.06 ± 0.015 $k_B$/vortex. In the layered compound YBCO values of ~0.4 $k_B$/vortex/layer are reported in comparable fields applied perpendicular to the superconducting layers [35]. It is a delicate question how to compare the latent heat of a vortex in an isotropic system as Nb$_3$Sn with that of a layered compound. A relevant length scale is the average length of vortex segments which are 'straight ' along the field direction [36] and it can be assumed that this length is closely related to the coherence length. We thus normalize the $\Delta S$-value by the coherence length ($\xi$ =36 Å for Nb$_3$Sn) which extends in Nb$_3$Sn over a few lattice parameters (5.293 Å) [12]. We find 0.3 ± 0.1 $k_B$/vortex/$\xi$, comparable to the value of YBCO and thus a reasonable value for a vortex-melting transition. A four times smaller value is derived from the AC specific-heat data in 6 T indicating an underestimation of the latent heat due to the presence of a temperature hysteresis and thus confirming the first-order nature of the transition.

In our interpretation the method of 'vortex shaking' seems to help restore the Bragg glass phase in the magnetization peak effect region where the thermal fluctuations soften the vortices and tend to transform the Bragg glass into a pinned intermediate glassy phase. Without vortex shaking, this glassy phase separates the Bragg glass and the liquid phase with no latent heat due to a first-order transition therefore being observed. Instead, a small hysteresis loop due to enhanced flux pinning appears in the magnetization (Fig. 2a inset). The solid-to-liquid transition has become a continuous crossover. Once the Bragg glass is restored by the AC field it melts directly into a liquid via a first-order melting transition.

In Fig. 5 we finally compare the 6 T data taken in presence of an AC field with the 3d-LLL fits [9] of the solid and liquid vortex phases identical to the curves used as a scaling function in Fig 3b. At the experimental melting temperature the difference in the two exactly represents the additional degrees of freedom due to the transition into the liquid represented by the 'idealized' step in the original measurement. In the vortex shaking experiment the step is hidden by the latent heat of the first-order vortex melting transition. We also include the metastable phases as dotted lines, with the supercooled liquid and overheated solid terminating at the spinodal point $T_{SP}^{clean}$ at $a_T$ = -5.5 [9,29] recently observed in the dynamics of several low-$T_c$ compounds [37]. The actual melting temperature ($T_m$) is shifted downwards due to quenched disorder and appears as a smeared spike somewhat below the theoretical value $T_m^{clean}$ [29].

## CONCLUSION

The present data show that fluctuations and vortex melting show up not only in HTSCs but are also present in low-$T_c$ superconductors. This is valuable information for the interpretation of universal features in the phase diagram of type-II superconductors, e.g. the broadening of the $H_{c2}$ line in fields [25,38] and the peak effect [13-18].

## ACKNOWLEDGMENTS

We acknowledge stimulating discussions with A. Schilling, T. Giamarchi, M.G. Adesso, R. Flükiger and H. Küpfer.

**Figure Captions:**

FIG. 1 a) Total specific heat of a single crystal of $Nb_3Sn$ in fields from 0 – 14 T measured by an AC technique. b) Details of the specific heat in small fields close to $T_c$ showing the evolution of a small fluctuation 'peak' superimposed on the specific-heat jump. Inset of Fig. 1 a) Comparison of the specific heat in 1 T measured with an AC and a DC technique.

FIG. 2 a) Magnetization as a function of applied field at four different temperatures. The straight lines serve as a guide to the eye to show the deviation from linear behavior due to the onset of fluctuations close to $T_c(H)$. The inset shows a sharp hysteresis loop ('magnetization peak effect') due to enhanced flux pinning at the lower onset of the fluctuation regime which develops at lower temperatures (here 14 K) [17,18]. b) Derivative of the magnetization curves showing a small fluctuation peak superimposed on the broadened jump at $T_c(H)$. Consistent units $J\ gat^{-1}T^{-1}=Am^2\ gat^{-1}$ and $J\ gat^{-1}K^{-1}$ are used for the magnetization and the specific heat respectively. Here 1 gat (gram-atom) is 1/4 mole.

FIG. 3 a) 3d-XY and b) 3d-LLL scaling of the specific-heat data after subtraction of the normal-state contribution. A theoretical fit [9] according to a 3d-LLL model is added to b) (see text for details). c) 3d-XY scaling of the derivative of the magnetization.

FIG.4 a) Specific-heat data in 3, 4, 5 and 6 T with a small AC field (~10 G, 1 kHz) superimposed parallel to the DC field (open circles) which helps the vortices to reach equilibrium in comparison with the data in absence of the AC field as shown in Fig 1. Arrows mark the vortex-melting temperatures ($T_m$). b) Difference in the data with and without AC field showing the anomalies related to vortex melting more clearly. Inset: First-order vortex-melting transition in the magnetocaloric effect $M_T(H) = (dQ/dB)_{T=const}$ (inverted for clarity) as measured by a DC technique during a field sweep experiment [17].

FIG.5 Specific-heat data in 6 T with a small AC field (~10 G, 1 kHz) superimposed parallel to the DC field (open circles) in comparison with theoretical fits (lines) [9] using a model for a solid vortex phase and a liquid (see text for details). $T_m$ is the first-order vortex melting transition, $T_m^{clean}$ the theoretical melting temperature in the absence of quenched disorder and $T_{SP}^{clean}$ the spinodal point.

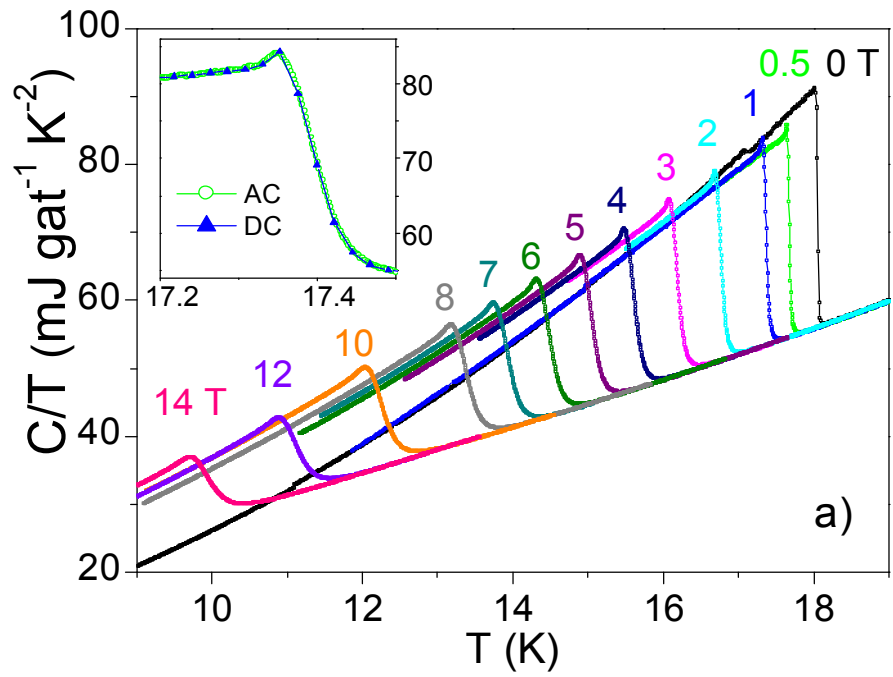

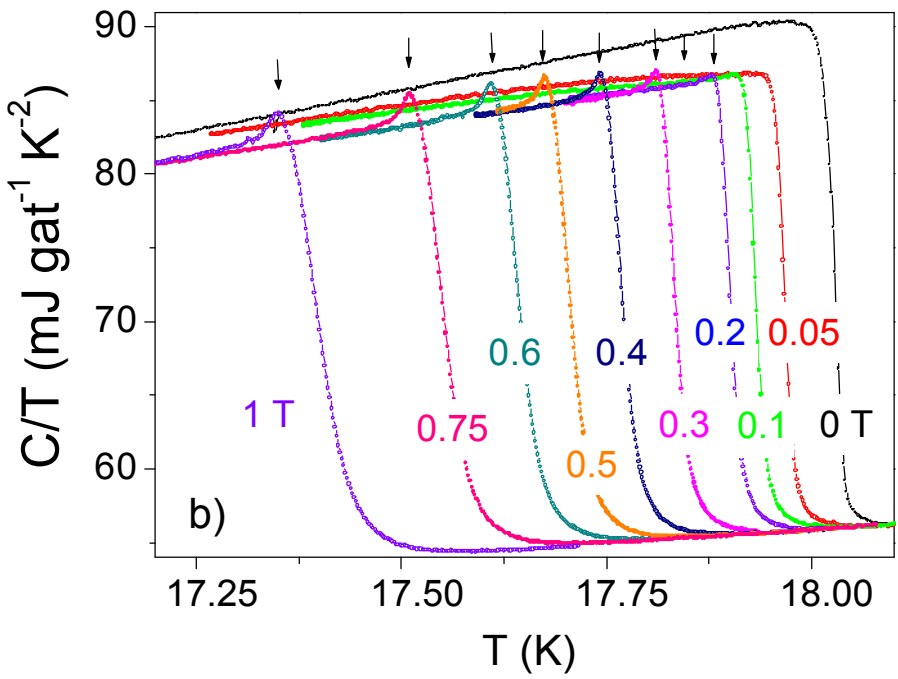

FIG 1a
FIG 1b

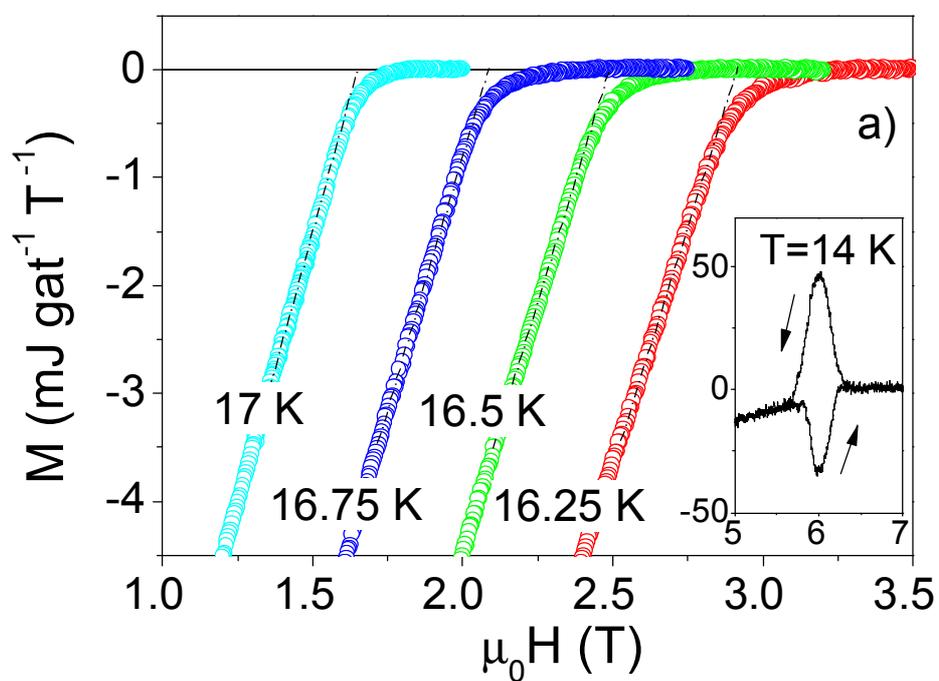
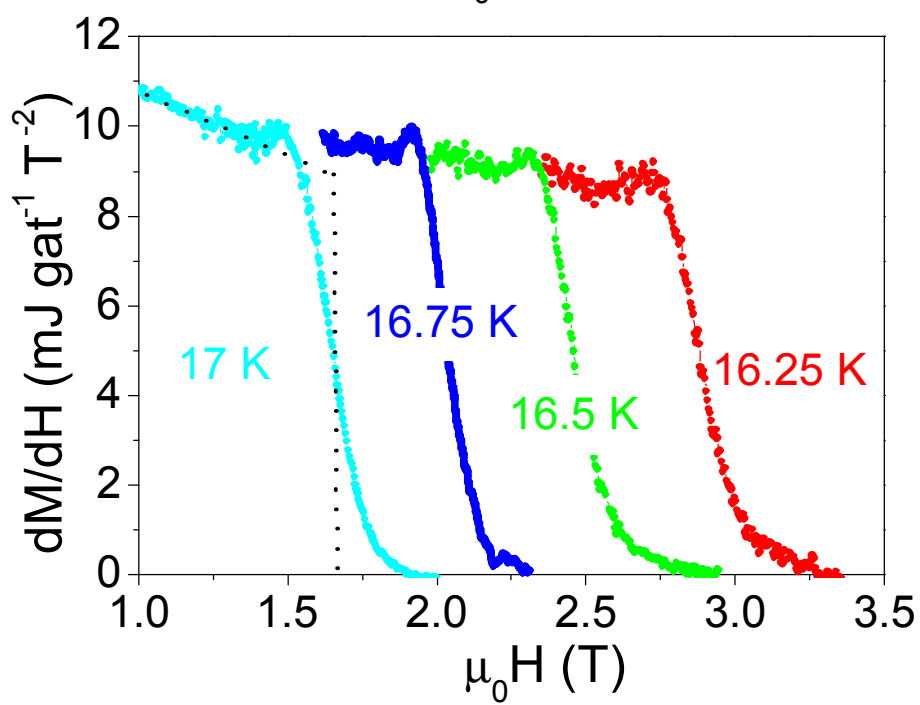

**FIG 2a**
**FIG 2b**

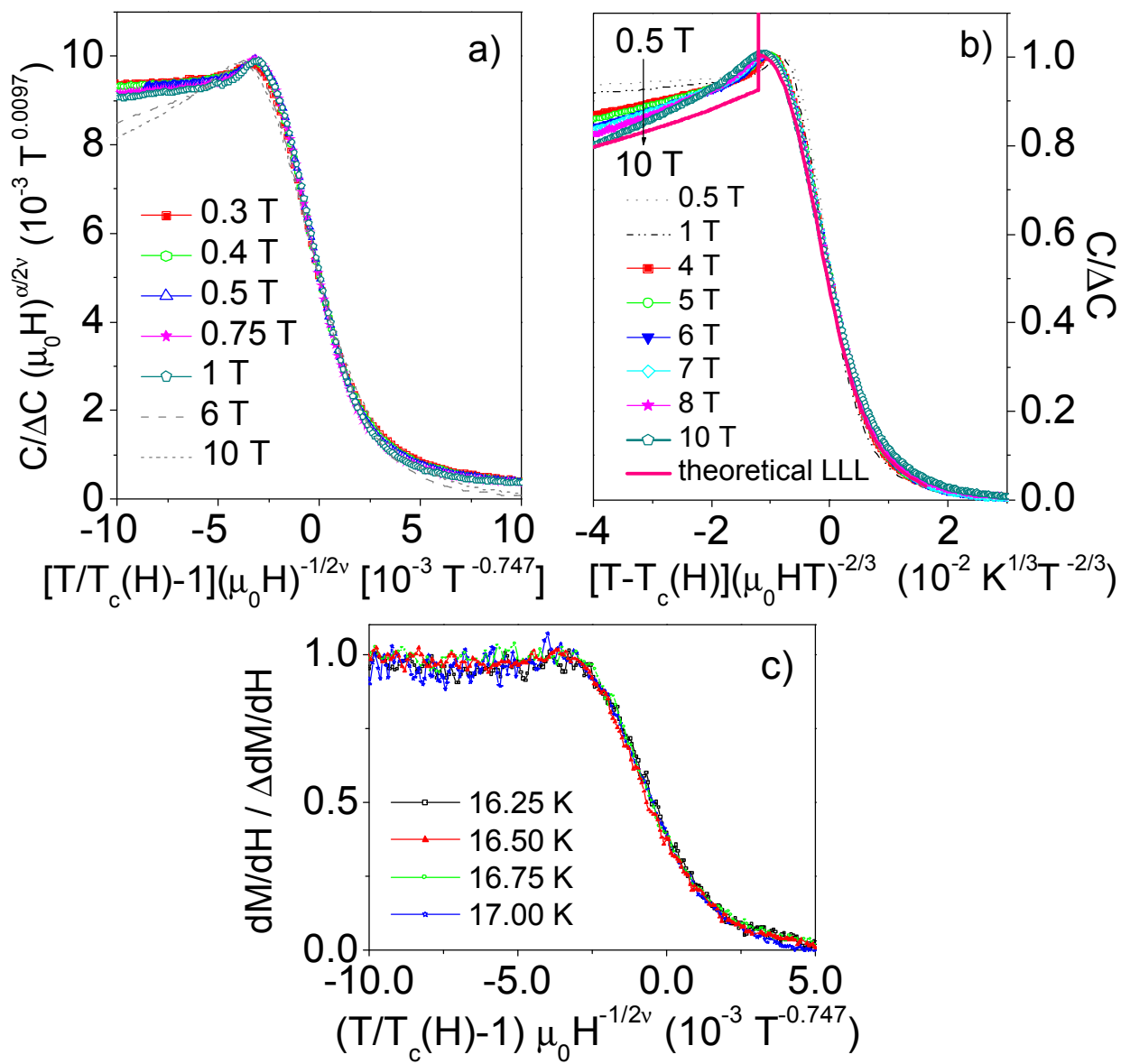

**FIG 3a**
**FIG 3b**
**FIG 3c**

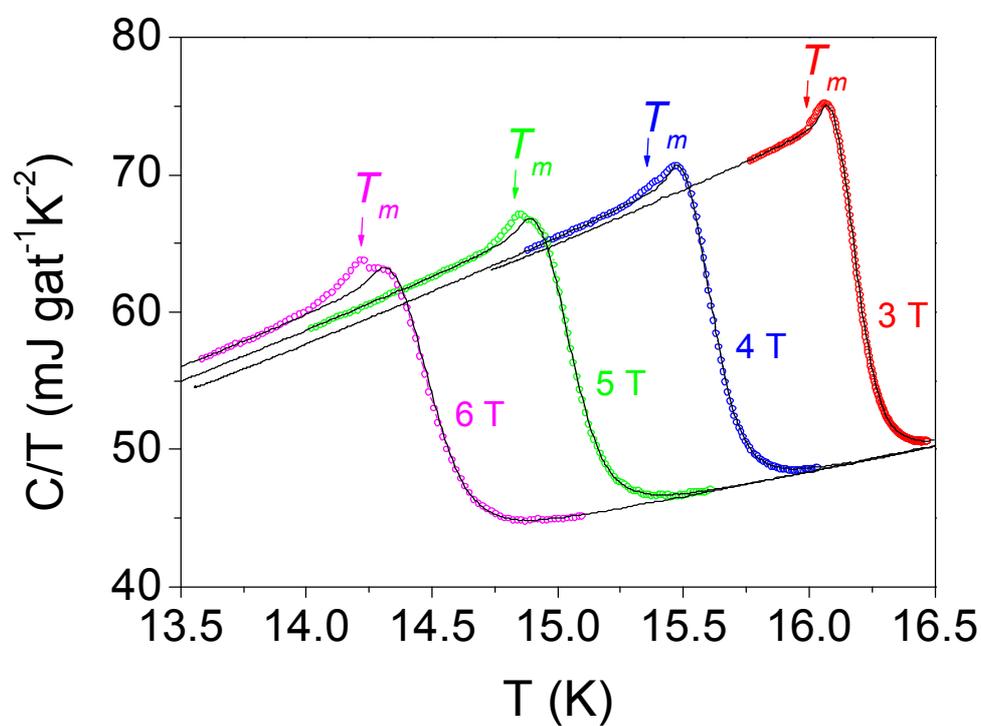

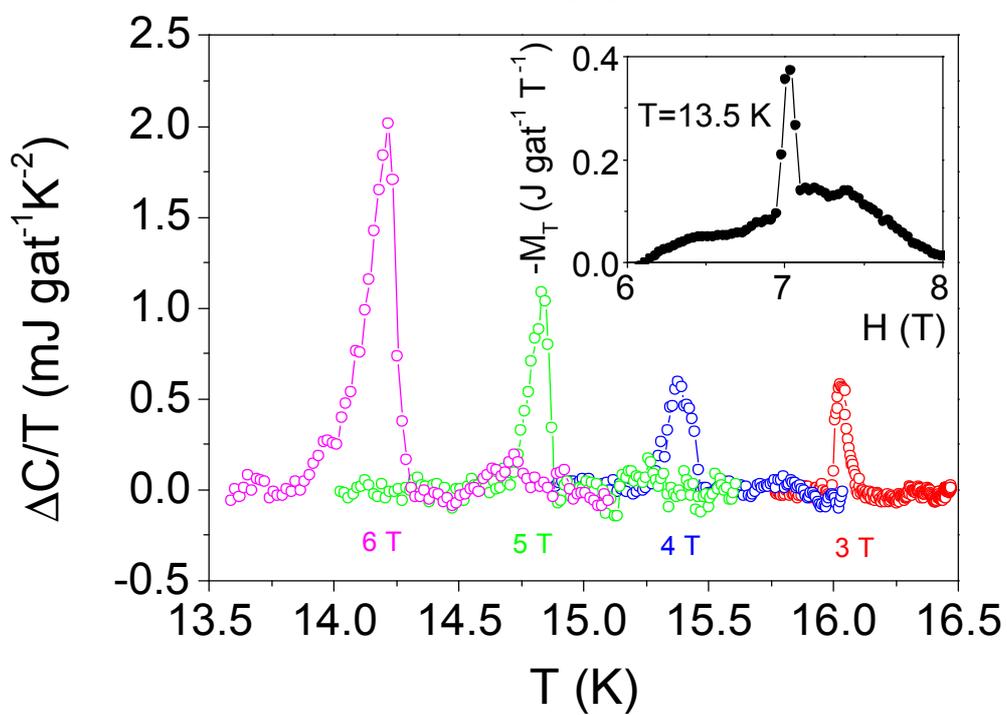

**FIG 4a**
**FIG 4b**

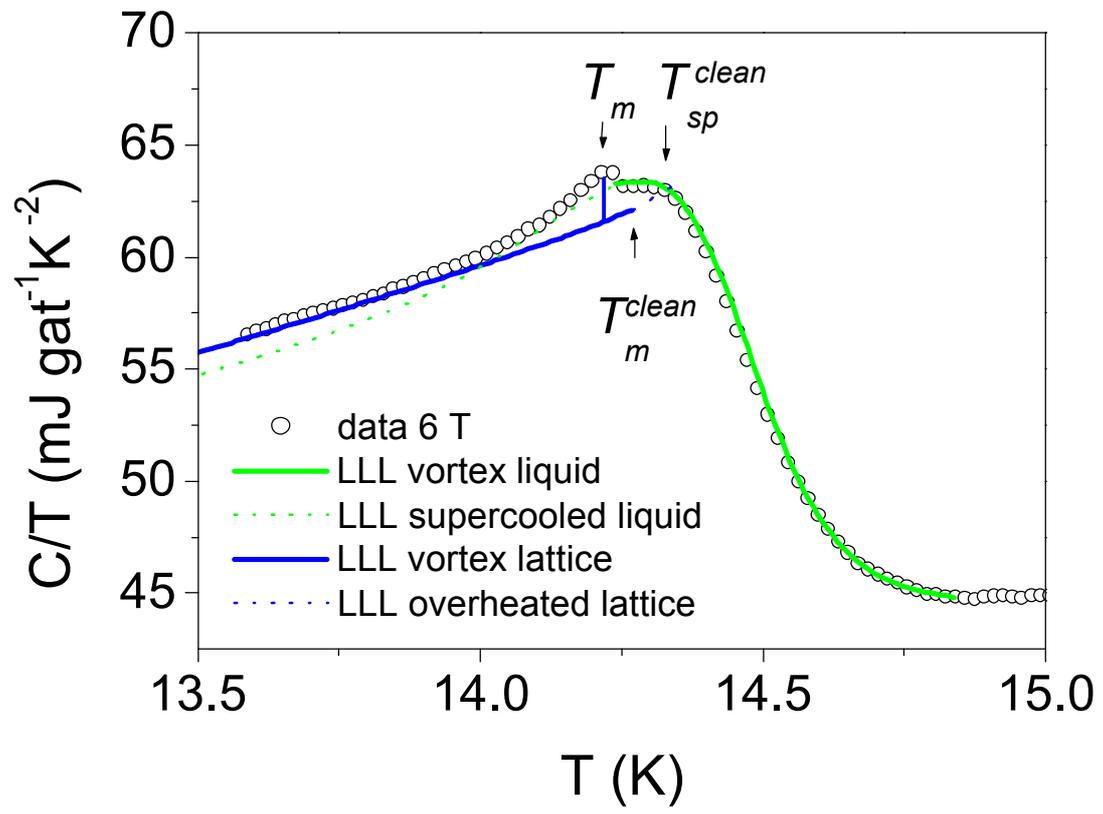

**FIG 5**